\documentclass[twocolumn,aps,prb,showpacs,widetext,superscriptaddress]{revtex4}
\usepackage{mathrsfs}
\usepackage{bm}
\usepackage{multirow}
\usepackage{amsmath,amsfonts,amssymb}
\usepackage{array,booktabs,fancyhdr}
\usepackage{graphicx,graphics,color,epsfig,times}

\begin{document}

\title{Band offset of GaAs/Al$_{x}$Ga$_{1-x}$As heterojunctions from atomistic first principles}
\author{Yin Wang}
\email{yinwang@hku.hk}
\affiliation{Department of Physics and the Center of Theoretical and Computational Physics, The University of Hong Kong, Pokfulam Road, Hong Kong SAR, China}
\author{Ferdows Zahid}
\affiliation{Department of Physics and the Center of Theoretical and Computational Physics, The University of Hong Kong, Pokfulam Road, Hong Kong SAR, China}
\author{Yu Zhu}
\affiliation{Nanoacademic Technologies, Brossard, PQ, Canada, J4Z 1A7}
\author{Lei Liu}
\affiliation{Nanoacademic Technologies, Brossard, PQ, Canada, J4Z 1A7}
\author{Jian Wang}
\affiliation{Department of Physics and the Center of Theoretical and Computational Physics, The University of Hong Kong, Pokfulam Road, Hong Kong SAR, China}
\author{Hong Guo}
\affiliation{Center for the Physics of Materials and Department of Physics, McGill University, Montreal, PQ, Canada, H3A 2T8}
\affiliation{Department of Physics and the Center of Theoretical and Computational Physics, The University of Hong Kong, Pokfulam Road, Hong Kong SAR, China}

\date{\today}
\begin{abstract}
Using an atomistic first principles approach, we investigate the band offset of the GaAs/Al$_x$Ga$_{1-x}$As heterojunctions for the entire range of the Al doping concentration $0<x\leq1$. We apply the coherent potential approach to handle the configuration average of Al doping and a recently proposed semi-local exchange potential to accurately determine the band gaps of the materials. The calculated band structures of the GaAs, AlAs crystals and band gaps of the Al$_x$Ga$_{1-x}$As alloys, are in very good agreement with the experimental results. We predict that valence band offset of the GaAs/Al$_x$Ga$_{1-x}$As heterojunction scales with the Al concentration $x$ in a linear fashion as $VBO(x)\simeq 0.587 x$, and the conduction band offset scales with $x$ in a nonlinear fashion.  Quantitative comparisons to the corresponding experimental data are made.
\end{abstract}
\maketitle

\emph{Introduction.} The properties of III-V compound semiconductors and their heterojunctions have been relentlessly investigated for several decades due to their wide-ranging applications in electronic and optoelectronic technologies. One of most important electronic property of heterojunctions is the band offset which describes the relative alignment of the electronic bands across the junction interface.\cite{Franciosi} An accurate determination of the band offset is critical for understanding quantum transport properties of the heterojuncton. For many III-V materials systems, the band offset has been carefully measured experimentally.\cite{Vurgaftman,batey}

Theoretical calculations of band offset have always been a serious challenge. This is because first principles method of density function theory (DFT) with local-density approximation (LDA)\cite{LDA} and generalized gradient approximation (GGA)\cite{GGA} underestimates the band gap ($E_g$) of semiconductors. Without a correct calculation of $E_g$ for individual semiconductors, the calculated band offset between two semiconductors may be compromised. Advanced methods such as GW\cite{GW} and/or hybrid functional\cite{HSE} can yield accurate $E_g$ for many systems but require significantly more expensive computation. Another difficulty is when there are impurities: the predicted physical results must be averaged over the multitudes of impurity configurations which is extremely costly in computation. Realistic semiconductors all have impurities and if the impurity concentration is small, one needs to compute systems of large number of host atoms in order to accommodate a few impurity atoms. Because of these issues, predicting band offset has persisted to be a challenging theoretical problem.

Considerable theoretical efforts have been devoted in the literature to correctly predict $E_g$. Apart from the GW\cite{GW} and hybrid functionals\cite{HSE}, for pure semiconductors the recently proposed modified Becke-Johnson (MBJ) semilocal exchange potential was shown to give quite accurate $E_g$ values for many compounds with a computational cost similar to that of LDA.\cite{MBJ} Even though MBJ is not a fundamental solution to the issue of electron correlation, it is practically useful for calculating $E_g$.  To deal with the prohibitively large computation required for performing configuration average of doped semiconductors, one wishes to obtain the averaged physical quantity without individually computing each impurity configuration by brute force as in the super-cell approach. In this regard, a widely used technique is the coherent potential approximation (CPA)\cite{CPA} as implemented in Korringa-Kohn-Rostoker\cite{KKR} or linear muffin-tin orbital (LMTO)\cite{LMTO} DFT methods. CPA is a statistical effective medium approach such that an atomic site has $x\%$ chance to be an impurity (e.g. dopant) atom and $(1-x)\%$ chance to be a host atom, the configuration average is carried out analytically hence disorder effect can be calculated for any concentration $x$. Very recently, Ref.~\onlinecite{GaN} has combined CPA with MBJ and reported the calculation of $E_g$ for the semiconductor In$_x$Ga$_{1-x}$N, the results are in excellent agreement with the measured data for the entire range of $0\leq x\leq 1$.

In this paper, we employ the CPA-MBJ approach to quantitatively calculate the band offsets of two semiconductors with impurity doping. In particular, we consider the most important heterojucntion, between GaAs and Al$_{x}$Ga$_{1-x}$As, and calculate the band offset as a function of the Al concentration $x$.  Even though this heterojunction has been the subject of extensive past investigations, theoretically the calculation was usually done at a few special values of $x$ where the super-cell approach could be applied. Here, the CPA-MBJ approach allows us to determine a continuous curve of the band offset for the entire range of $0< x \leq 1$, which has not been derived before from atomic first principles.  Our calculated $E_g$ of Al$_{x}$Ga$_{1-x}$As and the calculated band offsets of the heterojunctions for the entire $x$ range, are quantitatively and excellently compared with the corresponding experimental data.

\emph{Calculation details.} The lattice constants of the zinc-blende AlAs and GaAs were individually relaxed by a projector augmented wave method with LDA as implemented in the VASP software,\cite{VASP} and the optimized lattice constant of 5.63 {\AA} for GaAs were used for all the compounds in our calculations since the optimized lattice constant of AlAs (5.64 {\AA}) is very close to that of GaAs. To calculate the band offset with Al doping, we apply the CPA-MBJ approach as implemented in the Nanodsim software package\cite{nanodsim} where the DFT is carried out within the TB-LMTO scheme under the atomic sphere approximation (ASA).\cite{LMTO} For technical details of the Nanodsim algorithm we refer interested readers to Ref.~\onlinecite{youqi,nanodsim}.

In the DFT-CPA-MBJ calculations\cite{foot2}, the primitive cell of the zinc-blende structure was used to calculate the band structures and $E_g$. To determine the band offset, a (110) system containing 9 layers of GaAs and 9 layers of Al$_{x}$Ga$_{1-x}$As are used to calculate the potential profile through the heterojunction\cite{foot1}. A $12\times12\times12$ $k$-mesh and a $12\times12\times1$ $k$-mesh were used to sample the Brillouin Zone for the primitive cell (bulk) and the heterojunction, respectively. For the ASA, vacancy spheres were placed at appropriate locations\cite{GaN} for space filling, and the same radius were used for all the vacancy spheres and atomic spheres. Spin orbital coupling was not considered in this work.

\begin{figure}
\includegraphics[width=\columnwidth]{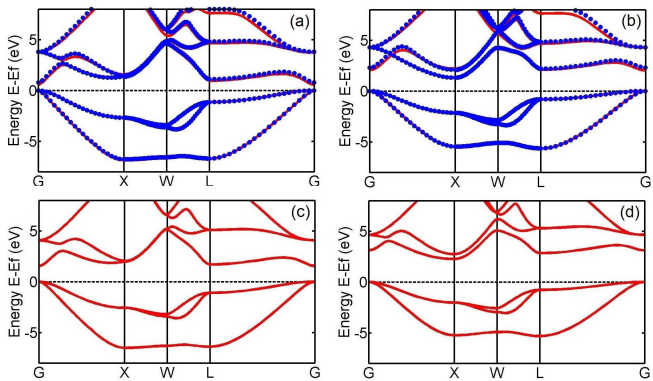}
\caption{(color online) (a,b) The band structures obtained with LDA:  (a) for GaAs, (b) for AlAs.  Red line is obtained by VASP, blue dots obtained by Nanodsim. (c,d) The band structures calculated with MBJ by Nanodsim: (c) for GaAs and (d) for AlAs.}
\label{fig1}
\end{figure}

A technical issue of MBJ semilocal exchange potential is worth mentioning. MBJ potential has the following form\cite{MBJ},
\begin{equation}\label{eq-MBJ}
v_{x,\sigma}^{MBJ}(r)=cv_{x,\sigma}^{BR}(r)+(3c-2)\frac{1}{\pi}\sqrt{\frac{5}{12}}
\sqrt{\frac{2t_{\sigma}(r)}{\rho_{\sigma} (r)}},
\end{equation}
where subscript $\sigma$ is spin index, $\rho_{\sigma}$ is the electron density, $t_{\sigma}$ is the kinetic energy density, and $v_{x,\sigma}^{BR}(r)$ is the Becke-Roussel potential.\cite{BR} The relative weight of the two terms is given by a parameter $c$ which depends linearly on the square root of $|\nabla \rho|/ \rho$. For all solids investigated by MBJ potential so far,\cite{MBJ} it appears that $E_g$ increases monotonically with $c$. The $c$ parameter can be determined self-consistently as discussed in Refs.~\onlinecite{MBJ,IMBJ}, but in this work we fixed its value to reproduce the experimental band gaps of GaAs, AlAs and, afterward, we used the same value to calculate the electronic structures of the alloy Al$_x$Ga$_{1-x}$As and the band offset of the heterojunctions.

\emph{Band structures of GaAs and AlAs.} We begin by calculating the band structures of GaAs and AlAs with LDA using both VASP\cite{VASP} and Nanodsim\cite{nanodsim} electronic packages. Fig.1 (a,b) show a perfect agreement of the valence bands and a good match of the conduction bands between these methods, confirming that the ASA employed in the TB-LMTO approach (Nanodsim) is accurate for calculating physical properties of these materials. It should be noted from Fig.1(a,b) that the band gaps were underestimated by LDA. Next, we apply the MBJ functional\cite{MBJ, IMBJ} to calculate the electronic structure again\cite{foot3} and the MBJ restuls are shown in Fig. 1(c,d). Compare with the LDA bands in Fig.1(a,b), the opening of band gap is evident. From Fig.1(c,d), the band gap values at $\Gamma$, $X$ and $L$ points are listed in Table~\ref{tab2}. The MBJ band gaps are in good agreement with the experimental values at the $\Gamma$ and $X$ points, and within 7\% for GaAs and 15\% for AlAs to the experimental values at the $L$ point. In contrast, the LDA gaps typically underestimate the values quite significantly which is a well known issue of LDA.

\begin{table}[ht]
\caption{Energies of the conduction band minima at the $\Gamma$, $X$, and $L$ points with respect to the valence band maximum at the $\Gamma$ point in units of eV, calculated by DFT with the $LDA$ and $MBJ$ functionals at zero temperature. The column of $LDA^v$ were obtained by VASP, other results were by Nanodsim within the TB-LMTO implementation. The last column are the experimental values.\cite{Vurgaftman}}
\centering
\begin{tabular} {ccccccc}
\hline\hline
Material$\ \ $ &$E_g\ \ $         &$LDA^v\ \ $ &$LDA\ \ $ &$MBJ\ \ $   & Expt.\cite{Vurgaftman}  \\ \hline
GaAs     &$\Gamma$ &0.493   &0.761 &1.518   &1.519                    \\
         &$X$      &1.334   &1.346 &1.960   &1.981                    \\
         &$L$      &0.948   &1.100 &1.691   &1.815                    \\ \hline
AlAs     &$\Gamma$ &2.014   &2.300 &3.099   &3.099                    \\
         &$X$      &1.312   &1.307 &2.258   &2.24                     \\
         &$L$      &2.086   &2.191 &2.835   &2.46                     \\ \hline\hline
\end{tabular}
\label{tab2}
\end{table}

\emph{Band gaps of Al$_x$Ga$_{1-x}$As.} Having accurately determined the band structures for the pure materials GaAs and AlAs with the MBJ functional, we apply the CPA-MBJ approach discussed above to calculate the the electronic properties of the alloy Al$_x$Ga$_{1-x}$As. The calculated band gaps are plotted versus the Al concentration $x$ in Fig.2. It is important to recall - as shown in Fig.1, that GaAs is a direct-gap semiconductor where the $\Gamma$ valley is lower than the $X$ valley in the conduction band; AlAs is an indirect-gap semiconductor where the opposite is true, namely $X$ valley is lower than the $\Gamma$ valley. By increasing the Al concentration $x$, the alloy changes from direct-gap to indirect-gap at a crossover point where the conduction band minima at $\Gamma$ and $X$ points have the same energy. The existence of the crossover point, at near $x \approx 0.36$, is quite evident in Fig.2. Such a crossover of band gap behavior is actually known experimentally and, as stated in Ref.~\onlinecite{Vurgaftman}, the experimental data implied the $\Gamma$-$X$ crossover composition to be at $x=0.38$ at low temperature and at $x=0.39$ at 300 K. It was also experimentally known that the $E_g$ scales linearly with $x$ in the direct-gap regime and quadratically in the indirect-gap regime.\cite{Vurgaftman, casey} Fitting the calculated results of $x\leq 0.3$ by a linear function and $x\geq 0.6$ by a quadratic function, as shown by the solid lines in Fig.2 excellent consistency to these scalings is obtained. In particular, the data points of $x=0.4$ and $x=0.5$ locate on the fitted curves and the crossover point can be seen at $x \approx 0.36$ which agrees reasonably well with the experimental observation of the crossover at $x=0.385\pm0.016$.\cite{Vurgaftman, cross}

Since the direct-to-indirect gap crossover is where the conduction band minima at $\Gamma$ and $X$ have the same energy value, it would be very useful to plot band structures of the  Al$_x$Ga$_{1-x}$As alloy. This is however not possible because when Al is randomly doped into GaAs, the material is disordered and momentum is no longer a good quantum number. Nevertheless, one can plot the calculated density of states (DOS) as a function of the momentum $\textbf{k}$ and energy $E$ as shown in the inset of Fig.2 which is for the alloy Al$_{0.36}$Ga$_{0.64}$As. This inset reveals a broadened ``band structure" of the disordered alloy where the ``bands" are no longer infinitely sharp lines as that of crystals due to impurity broadening. The broadened ``bands" trace out the energy minima and dispersion from which we found that the ``conduction band" minima at $\Gamma$, $X$, and $L$ points have essentially the same energy value for this alloy, thus confirming that the theoretical crossover point is at $x \approx 0.36$.

\begin{figure}
\includegraphics[width=\columnwidth]{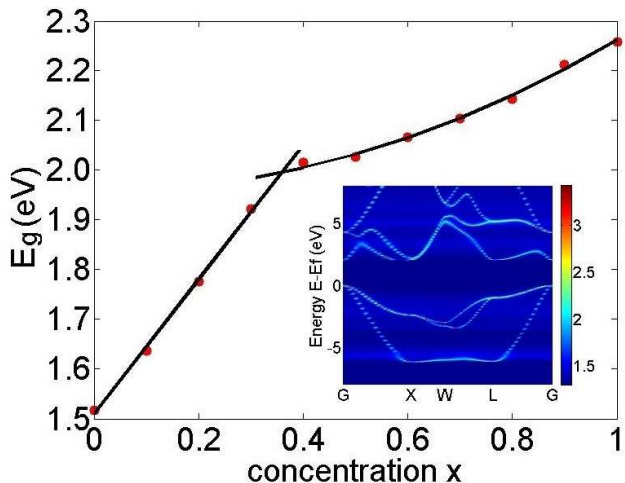}\\
\caption{(color online) The calculated band gaps of Al$_x$Ga$_{1-x}$As versus $x$ by the CPA-MBJ approach. The two solid lines are fitting to the data in the two ranges of $x$. Inset: the calculated DOS for the alloy Al$_{0.36}$Ga$_{0.64}$As in logarithmic scale as a function of momentum $k$ and energy $E$, revealing a broadened ``band structure". }
\label{fig2}
\end{figure}

\begin{figure}
\includegraphics[width=\columnwidth]{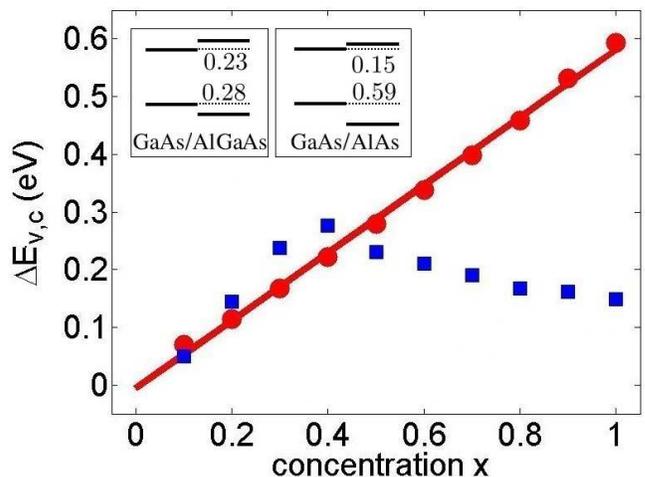}\\
\caption{(color online) Valence band offset (red dot) and conduction band offset (blue square) at different concentration x. The red line shows the linear fitting of the VBO. The GaAs/Al$_x$Ga$_{1-x}$As heterojunctions have the straddling type gap - the valence band maximum (VBM) of GaAs is higher, while its conduction band minimum (CBM) is lower. Insets: band alignment diagrams of GaAs/Al$_{0.5}$Ga$_{0.5}$As (on the left) and GaAs/AlAs (right). The VBO (CBO) values are with respect to the VBM (CBM) of GaAs in units of eV.}
\label{fig3}
\end{figure}

\emph{Band offsets of heterojunctions.} Having correctly determined the band gaps, dispersions and the direct-to-indirect gap crossover for the alloy Al$_x$Ga$_{1-x}$As, the band offset of the GaAs/AlGaAs heterojunction can be confidently analysed. The valence band offset (VBO) and the conduction band offset (CBO) of a heterojunction are defined as the difference between the energy values of the top of the valence bands and the bottom of the conduction bands of the two materials forming the junction, respectively. As discussed in the introduction, band offsets are of great importance for transport properties of the heterojunction. The band offsets can be calculated as:
\begin{equation}
  VBO,\  CBO = \Delta E_{v,c} + \Delta V,
\end{equation}
where $\Delta E_v (\Delta E_c)$ is defined as the difference between the top (bottom) of the valence (conduction) bands of two independent bulk materials that form the heterojunction, $\Delta V$ is the lineup of the potential through the heterojunction. When the two materials on either side of the heterojunction is extended long enough, $\Delta V$ is independent of the orientation of the two material interfaces.

Previously, by supercell method the VBO of the GaAs/Al$_x$Ga$_{1-x}$As heterojunction were calculated at a few particular values of $x$, and several different scalings of VBO versus $x$ were reported. For instance Ref.~\onlinecite{wang} reported VBO $\sim0.564x-0.032x^2$ (units eV);  Ref.~\onlinecite{ribeiro} reported VBO $\sim0.41x^2$; and Ref.~\onlinecite{ekpunobi} reported VBO $\sim0.17x$. Recently, Ref.~\onlinecite{wadehra} reported a calculation at $x=0.5$ by the hybrid density functional method and obtained VBO=0.26 and CBO=0.42. However, first principles analysis of band offsets for the entire $0<x\leq 1$ range is still lacking.

By the CPA-MBJ method discussed above, we have calculated VBO and CBO of GaAs/Al$_x$Ga$_{1-x}$As heterojunction for the entire Al concentration range, results plotted in Fig.~\ref{fig3} versus $x$ in intervals of $0.1$. The two insets of Fig.\ref{fig3} plot the band alignment diagrams for $x=0.5$ and $1.0$. It is apparent that VBO is a linear function of $x$ while CBO is not. By fitting VBO linearly (solid red line) we obtain $VBO(x)\simeq 0.587 x$ eV which agrees quite reasonably with the experimental scaling\cite{batey} of $VBO(x)\simeq0.55 x$ eV. The calculated VBO values in Fig.~\ref{fig3} agree very well with the previous experimental and theoretical reports,\cite{Vurgaftman,christensen} namely at $x=0.3$, the calculated VBO is 0.167 eV, to be compared with the experimental value of $0.17\pm0.04$ eV.\cite{Gwo} Concerning CBO, it was known experimentally that CBO appears to fit an approximate relation CBO$\simeq 0.6 \Delta E_g^\Gamma$ where $\Delta E_g^\Gamma$ is the band gap difference of the two parts of the heterojunction.\cite{batey, miller} The linear increase of the CBO in the direct gap range in Fig.~\ref{fig3} is very close to this relation. Near and beyond the direct gap range, the calculated CBO decreases with increasing of $x$, which is also in good agreement with the experimental results.\cite{batey}

\emph{Summary.} Using the DFT-CPA-MBJ first principles approach we have calculated the band offset of the GaAs/Al$_x$Ga$_{1-x}$As heterojunctions for the entire range of the Al doping concentration $x$. The calculated offset scales with the Al concentration $x$ in a linear fashion as $VBO(x)\simeq 0.587 x$, and the conduction band offset scales with $x$ in a nonlinear fashion.
In our calculations, the impurity doping is handled by the CPA while accurate band structures and band gaps of the materials are obtained by the MBJ semilocal exchange potential. Our atomistic calculations of the band structures of GaAs and AlAs crystals, band gaps of the Al$_x$Ga$_{1-x}$As alloy, and band offsets of GaAs/Al$_x$Ga$_{1-x}$As heterojunctions compare very well with the corresponding experimental results. From the calculated density of states, a broadened ``band structure" at the direct-to-indirect gap crossover point is obtained at which the alignment of the conduction band minima at the $\Gamma$, $X$ and $L$ points occur, also in agreement to the experimental observation.

We thank Dr. J.N. Zhuang, Dr. B. Wang, Dr. H.T. Yin, and Mr. R.G. Cao for useful discussions, and CLUMEQ for providing computation facilities. This work is supported by the University Grant Council (Contract No. AoE/P-04/08) of the Government of HKSAR and NSERC of Canada (H.G).


\begin{thebibliography}{100}
\bibitem{foot2}
During the self-consistent DFT-CPA-MBJ iterations, Ga and As $3d$  electrons are treated as core electrons and the $4d$ orbitals are included in the self-consistent process.
\bibitem{foot3}
The optimized $c$ parameters used in the MBJ functional for calculating the GaAs and AlAs materials are: (a) For GaAs crystal, $c=1.09$ for the atoms and $1.34$ for the vacancy spheres. (b) For AlAs crystal, $c=1.04$ for the atoms and $1.44$ for the vacancy spheres. These values are used for all other calculations presented in this work.
\end{thebibliography}
\end{document}